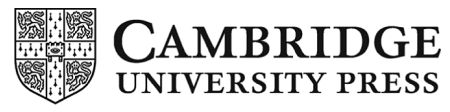

ARTICLE

# Great Disappearance Acts — Generative Search and Shadow Banning

Danny Friedmann

Peking University School of Transnational Law, China
Email: dannyfriedmann@stl.pku.edu.cn

**Abstract**

The internet, once celebrated as a decentralized public sphere, is increasingly undermined by practices such as generative search and shadow banning, which divert traffic and suppress visibility. Generative search, powered by Retrieval-Augmented Generation (RAG), synthesizes content into direct answers, bypassing websites and depriving them of traffic and revenue. This threatens the sustainability of independent content creators, small enterprises, and the open web ecosystem. Shadow banning, a practice that intentionally reduces the visibility of social media posts through algorithmic moderation, exacerbates these issues by chilling free expression and limiting transparency and accountability.

This article explores these opaque practices through a legal and regulatory lens. The first part examines the rise of generative search, analyzing its technological and legal implications, including copyright infringement, unfair competition, and unjust enrichment. It also evaluates potential solutions such as licensing agreements and agentic AI. The second part focuses on shadow banning: algorithmic dissuasion, de-ranking, and the reduction of traffic, with specific attention to China's Regulation on Algorithmic Recommendations (RAR) and the EU's Artificial Intelligence Act (AIA). Both frameworks offer partial solutions but fall short of ensuring fairness, transparency, and redress mechanisms.

Ultimately, the shift toward centralized control by dominant platforms prioritizes profit and risk management over innovation, fairness, and diversity in online expression. To counteract these trends, regulatory interventions, algorithmic transparency, and equitable frameworks are essential. Without such measures, the internet risks losing its character as a democratized public sphere for free expression and innovation.

## A. Introduction

There are two opaque ways in which the internet's ecosystem—encompassing commercial websites owned by small- and medium-sized enterprises, citizen journalists, weblogs, and social media platforms—is being undermined by the diversion of traffic, the lifeblood of digital publication, away from its original content.

At least at first sight, generative search results unintentionally do this, while shadow banning does so intentionally. This is of great importance, because websites and social media posts are the modern version of the public sphere, decentralized spaces where individuals can share news, or expressive information, and make a living, which should be able to counterbalance to centralized commercial news and commodified information. The problem will be exacerbated by a deluge of synthetic content on websites[1] and social media posts, which will dilute further the human-created content.

Jürgen Habermas described how the public sphere emerged, separate from the private and state spheres, in the 17th and 18th centuries.[2] Mercantilism and early forms of capitalism contributed to the rise of the bourgeois class, which, in turn, played a critical role in fostering the Enlightenment and the development of the bourgeois public sphere.[3] Within this sphere, salons and coffee houses served as key institutions where members of the bourgeoisie gathered to engage in rational-critical debate about politics, culture, and society.[4] Mass media was becoming commodified and transformed journalism into a market-driven service[5] prone to manipulation,[6] causing an erosion of the true public debate[7] and changing the democratic authenticity of public opinion. This was further aggravated by the concentration of media ownership.[8]

In contrast, the advent of the internet, particularly weblogs[9] and later social media,[10] has been embraced as a means to democratize news and provide new opportunities for self-expression. One could argue that these websites have evolved into "the modern public square," as the Supreme Court described social media in 2017 in *Packingham v. North Carolina*,[11] constituting "the most dynamic forum for speech in the United States."[12] Anyone with a computer and an internet connection can create a website and share their thoughts, experiences, news, knowledge, and expertise at minimal cost, or earn a living. Yet today, generative search and shadow banning threaten to replicate and intensify the very dynamics that Habermas critiqued. Generative search abstracts and repackages the expressive contributions of individual creators, severing the link between speaker and audience, while shadow banning quietly suppresses visibility without procedural transparency. Both practices displace the deliberative public from its own communicative outputs, filtering and fragmenting discourse in ways that privatize attention and algorithmically gatekeep participation. Generative search, which induces internet users "to skip the links," significantly decreases the traffic of visitors to the websites.[13] No visitors means no

[1] Wes McDowell, *Website Traffic Will Disappear in 2025. Prepare Now.*, (YouTube, Dec. 17, 2024), https://www.youtube.com/watch?v=db4tr3BAWJo.

[2] Jürgen Habermas, The Structural Transformation of The Public Sphere: An Inquiry into a Category of Bourgeois Society 14 (Thomas Burger & Frederick Lawrence trans., MIT Press 1989) (1962)

[3] *Id.* at 14, 17.

[4] *Id.* at 31.

[5] *Id.* at 181.

[6] *Id.* at 196.

[7] *Id.* at 211.

[8] *Id.* at 181.

[9] Neil McIntosh, *Google Buys Blogger Web Service*, The Guardian (Feb. 18, 2003), https://www.theguardian.com/business/2003/feb/18/digitalmedia.citynews (noting that Pyra Labs launched Blogger in 1999 and Google acquired it in 2003).

[10] William Fisher, *Freedom of Expression on the Internet*, Berkman Ctr. for Internet & Soc'y (June 14, 2001), https://cyber.harvard.edu/ilaw/Speech/.

[11] Packingham v. North Carolina, 582 U.S. 98, 107 (2017).

[12] *Id.*

[13] Yuval Halevi, *The Internet's Most-Read Tech Publications Have Lost 58% of Their Google Traffic Since 2024*, Growtika (Feb. 2026), https://growtika.com/blog/tech-media-collapse (noting that ten major tech publications lost between 30 and 90 percent: "For most of these publications, the traffic curves held through early 2025 and then dropped sharply in the second half of the year."). Gabriele Giuntini, *Google and AI Overview. The Collapse of Organic Traffic*, HT&T Cons. (Aug. 1, 2025), https://www.htt.it/en/google-and-ai-overview-the-collapse-of-organic-traffic/ (this consultancy company estimates a decline in organic traffic ranging from 18 to over 30 percent). Klaudia Jaźwińska, *Traffic Apocalypse: Google's AI Overviews Are Killing Click-Throughs to News Sites*, Colum. J. Rev. (July 31, 2025), https://www.cjr.org/analysis/traffic-apocalypse-google-ai-overviews-killing-click-throughs-news-sites.php (the estimate for a worldwide decrease in click-through rates over a year is 15 percent). *See also* Priya S., *Sudden Drop in Search Ranking and Traffic After Google March 2024 Core Update – Need Help*, Google Search Cent. Cmty. (Mar. 13, 2024), https://support.google.com/webmasters/thread/263667593/sudden-drop-in-search-ranking-and-traffic-after-google-march-2024-core-update-need-help?hl=en (claiming to have lost 90 percent of organic traffic in one day).

revenue stream based on advertisement, subscription, patronage, or crowdfunding. This is undermining the sustainability of these business models for websites, the very foundational real estate of the internet. This Article contends that these practices are not mere technical evolutions but structural threats that demand doctrinal, policy-based, and normative engagement. This Article interrogates how generative search and shadow banning—two opaque, algorithmically mediated practices—are reshaping the information architecture of the internet in ways that threaten the visibility of lawful expression, the sustainability of independent content creation, and the integrity of the digital public sphere. By analyzing these phenomena through the lenses of intellectual property, platform governance, and comparative regulatory frameworks, the Article aims to develop a normative and doctrinal response to the vanishing of source attribution and audience access in an AI-dominated search environment. It argues for a recalibration of copyright, tort, and administrative law to restore transparency, fairness, and pluralism in the evolving informational ecosystem.

This Article unfolds in two parts, mapping how generative technologies and algorithmic moderation disrupt the digital information ecosystem.

The first part, Sections B to D, deals with the impact of generative search, particularly the use of Retrieval-Augmented Generation (RAG), on the visibility and sustainability of web-based content. Section B, "From Riches to RAG," charts the internet's transformation from a decentralized network of independent content creators to a more centralized ecosystem dominated by a limited number of platforms wielding control over traffic flows and content distribution. Subsection I traces the historical trajectory of search engines, culminating in the rise of RAG systems; Subsection II focuses exclusively on the RAG framework itself; and Subsection III examines generative search as a salient subtype of RAG. Section C addresses the responses available to website holders confronting the disruptive effects of generative search. Subsection I outlines technological measures, including opt-out mechanisms; Subsection II examines the practical limitations and barriers to effective opt-out; and Subsection III analyzes *Dow Jones & Company v. Perplexity AI*, offering further legal avenues of redress: (a) copyright infringement; (b) plagiarism and semi-plagiarism as normative and reputational tools; (c) unfair competition; (d) unjust enrichment and negligence; (e) the potential of licensing arrangements; and (f) the prospective role of agentic AI in mitigation. Section D offers a synthesis of the first part's findings.

The second part, Sections E to G, turns to the phenomenon of shadow banning and its destabilizing effects on the digital public sphere through covert modulation of social media visibility. Section E introduces the conceptual contours of shadow banning. In the absence of comparable U.S. regulation, Subsection I analyzes China's and the European Union's frameworks governing algorithmic dissuasion, while Subsection II distinguishes between content de-ranking and traffic reduction. Section F focuses on the algorithmic suppression of content potentially implicated in copyright disputes. Subsection I explores the actors and consequences involved in such de-ranking, including: (a) measures to constrain recommendation algorithms; and (b) the treatment of unauthorized yet lawful user-generated content. Subsection II assesses the relevance of China's Regulation on Algorithmic Recommendations (RAR), and Subsection III evaluates the approach of the EU's Artificial Intelligence Act (AIA) to shadow banning practices. Section G concludes the second part. Section H provides the overall conclusion of the Article.

"Don't bite the hand that feeds you."[14] This saying warns against harming the source of your benefits. Applied to generative search using RAG, it reflects the tension between these systems and the websites they rely on for content. While RAG-powered generative search use websites as their foundation, they reduce traffic by offering direct, synthesized answers, depriving sites of ad revenue, subscriptions, and visibility.

[14]Edmund Burke, Thoughts on the Cause of the Present Discontents 3 (1770), Univ. of Mich. Libr. Digit. Collections, https://name.umdl.umich.edu/004903072.0001.000 (paraphrasing).

## B. Riches to RAG

This section examines the devolution of the ecosystem of the internet from predominantly a decentralized system, where control and content were distributed across many independent servers and networks, to a centralized system, with a few dominant platforms and corporations controlling significant portions of traffic, content, and user interactions.

### *I. Historical Development in Search Technology*

The first internet search engine, "Archie," was created in 1989.[15] It indexed downloadable files across the internet. In 1991 "Jughead" followed the Gopher protocol,[16] but searched a single server at a time.[17] In 1992, "Veronica" was used to browse and index information in Gopher menu items.[18] In 1994, the search engines "WebCrawler"[19]—and in 1995, "Lycos"[20]—allowed users to search the full text of web pages. In 1994, *Yahoo!* began as a web directory, manually cataloging websites, organizing content by categories to automate the indexing process. In 1995, "AltaVista"[21] introduced automatic indexing techniques and allowed users to search using natural language.

Google introduced its PageRank algorithm in 1998. It ranked websites based on the number and quality of links pointing to them.[22] Google constantly tweaked its search algorithm to improve the quality of the search and to increase AdWord pay-per-click advertising results. In 2013, Google released "Hummingbird," which signaled the beginning of semantic search.[23] Hummingbird focuses more on context than keywords alone. In 2015, "BrainRank"[24] was created. BrainRank takes factors such as the location of the searcher, personalization based on click-through rates, and word combinations into account, to better discern the intention of the user over time.

In 2018, Google enhanced search with "BERT," which stands for Bidirectional Encoder Representations from Transformers, to understand the nuances of language, especially word order and prepositions.[25] In 2021, Google's "MUM" (Multitask Unified Model) shifted search toward multimodal generative results, to understand language better by integrating information from different formats.[26]

---

[15]Kevin Purdy, *Archie, the Internet's First Search Engine, is Rescued and Running*, ARS TECHNICA (May 17, 2024), https://arstechnica.com/gadgets/2024/05/archie-the-internets-first-search-engine-is-rescued-and-running/.

[16]Scott Carlson, *How Gopher Nearly Won the Internet*, CHRONICLE (Sep. 5, 2016), https://www.chronicle.com/article/how-gopher-nearly-won-the-internet/ ("Gopher is a communication protocol designed for distributing, searching, and retrieving documents over the Internet."). *See also Archie, the first Internet search engine*, STACKSCALE GRUPO AIRE (Sep. 10, 2021), https://www.stackscale.com/blog/archie-internet-search-engine/ ("Both Jughead and Veronica were search engine systems for Gopher, a protocol developed by Mark McCahill in 1991, at the University of Minnesota.").

[17]Tom Seymour, Dean Frantsvog & Satheesh Kumar, *History of Search Engines*, 15 INT'L J. MGMT. & INFO. SYS. 47, 49 (2011), https://scispace.com/pdf/history-of-search-engines-1rl4z2qbz6.pdf ("Jughead searched a single server at a time.").

[18]*Veronica Search Engine*, WEB DESIGN MUSEUM, https://www.webdesignmuseum.org/web-design-history/veronica-search-engine-1992 (last visited Jan. 14, 2025).

[19]Owen Whitcombe, *The History of Search Engines,* LIBERTY MKTG. (May 26, 2022), https://www.libertymarketing.co.uk/blog/a-history-of-search-engines/.

[20]*Company Overview*, LYCOS, https://info.lycos.com/about/company-overview/ (last visited Jan. 14, 2025).

[21]Ernie Smith, *Whatever Happened to AltaVista, Our First Good Search Engine*, VICE (Jan. 4, 2021), https://www.vice.com/en/article/whatever-happened-to-altavista-our-first-good-search-engine/.

[22]Patrick Stox, *The Evolution of PageRank*, AHREFS BLOG (June 27, 2024), https://ahrefs.com/blog/google-pagerank/.

[23]Greg Kumparak, *Google Recently Made a Silent Shift to a New Search Algorithm, "Hummingbird,"* TECHCRUNCH (Sep. 26, 2013), https://techcrunch.com/2013/09/26/google-recently-made-a-silent-shift-to-a-new-search-algorithm-hummingbird/.

[24]John Rampton, *Artificial Intelligence is Changing SEO Faster Than You Think*, TECHCRUNCH (June 4, 2016), https://techcrunch.com/2016/06/04/artificial-intelligence-is-changing-seo-faster-than-you-think/.

[25]Jacob Devlin, Ming-Wei Chang, Kenton Lee & Kristina Toutanova, *BERT: Pre-training of Deep Bidirectional Transformers for Language Understanding*, ARXIV (May 24, 2019), https://arxiv.org/abs/1810.04805.

[26]Pandu Nayak, *MUM: A New AI Milestone for Understanding Information*, GOOGLE BLOG (May 18, 2021), https://blog.google/products/search/introducing-mum/.

### *II. Retrieval-Augmented Generation (RAG)*

In 2017, Anish Vaswani and others wrote "Attention Is All You Need."[27] This Article introduced the transformer model, which has become the foundation for Natural Language Processing (NLP) and deep-learning architectures, including Large Language Models (LLMs). The LLMs have made the generative AI revolution possible.

Traditional search engines, based on extractive systems, provide a list of relevant links, while AI services, based on abstractive systems, provide synthesized results. The downside of these generated answers is that LLMs have a tendency to hallucinate, leak information, and present wrong answers with the same aplomb as a right answer.[28] Moreover, LLMs do not provide any source;[29] after training, they cannot be updated or expanded in a cost-efficient way.[30]

On May 22, 2020, a consortium of researchers from Facebook AI Research, University College London, and New York University submitted the article "Retrieval-Augmented Generation for Knowledge-Intensive NLP Tasks."[31] This foundational article proposed a Retrieval-Augmented Generation (RAG) system that combines retrieval and generation,[32] to avoid the above-mentioned problems of LLMs,[33] at least to a certain extent.[34] The answers of a RAG system are grounded to the content in the sources of an external database to which it has access, and subsequently the LLM synthesizes this information in a representable and presentable form. To illustrate this system: If a user enters the prompt in a RAG system: "What are the causes of climate change?" In the semantic search, the vector of that prompt will be compared to the vectors of the external documents and the most similar and thus relevant documents or snippets will be retrieved. For example: "A scientific article on greenhouse gases;" "A news article about deforestation;" and "A government report on industrial emissions."

In the enhanced context phase, the system filters out less relevant information and formats the input for example as: "The following are the main causes of climate change based on retrieved documents: greenhouse gas emissions, deforestation, and industrial activities. Use this information to answer the query." After the semantic search and the enhanced context, the LLM is able to generate a coherent response: "Climate change is primarily caused by the accumulation of greenhouse gases, deforestation, and industrial emissions, which increase the Earth's temperature."

---

[27]Ashish Vaswani, Noam Shazeer, Niki Parmar, Jakob Uszkoreit, Llion Jones, Aidan N. Gomez, Lukasz Kaiser & Illia Polosukhin, *Attention Is All You Need*, ARXIV (June 12, 2017), https://arxiv.org/abs/1706.03762.

[28]IBM TECHNOLOGY, *What is Retrieval-Augmented Generation (RAG)?*, (YouTube, Aug. 23, 2023), https://www.youtube.com/watch?v=T-D1OfcDW1M.

[29]*Id.*

[30]*Id. See also* Eliza Mik, *Caveat Lector: Large Language Models in Legal Practice*, 19 RUTGERS BUS. L. REV. 70, 119 (2024) ("Their parametric knowledge not only abounds in incorrect information but becomes, as a matter of principle, fixed at a certain point in time.").

[31]*See generally* Patrick Lewis et al., *Retrieval-Augmented Generation for Knowledge-Intensive NLP Tasks*, ARXIV (May 22, 2020), https://arxiv.org/abs/2005.11401.

[32]Peter Henderson, Tatsunori Hashimoto & Mark Lemley, *Where's the Liability in Harmful AI Speech?*, 3 J. FREE SPEECH L. 589, 609 (2023) ("An intermediate approach between purely extractive systems and abstractive systems is called retrieval-augmented generation.").

[33]Nicholas Mignanelli, *The Legal Tech Bro Blues: Generative AI, Legal Indeterminacy, and the Future of Legal Research and Writing*, 8 GEO. L. TECH. REV. 298, 312 (2024) ("[L]imiting the length of interactions can mitigate the risk of 'hallucination,' [but] it will never be zero."). *See also* Marina Danilevsky, *What Is Retrieval-Augmented Generation (RAG?)*, at 5:21 (YouTube, Aug. 23, 2023) (uploaded by IBM Technology), https://www.youtube.com/watch?v=T-D1OfcDW1M ("If the user's answer cannot be reliably answered based on the data store, the model should say I do not know.").

[34]Complaint at 29–30, Dow Jones & Co. v. Perplexity AI, Inc., 797 F. Supp. 3d 205 (S.D.N.Y. 2025) (No. 24-cv-07984) (arguing (Dow Jones & Co.) that Perplexity AI's RAG system also produces hallucinatory results) [hereinafter Dow Jones Complaint].

The RAG programmer inserts a system-level instruction—effectively a prompt that precedes the user's prompt[35]—directing the model to prioritize this information. The user's query (prompt) is converted into a vector, which is a numerical representation.[36] The vector captures the semantic meaning of the query, rather than just its surface-level keywords.[37] After a user's query, the vector of the query is then compared to the vectors of the content in the database to identify which parts of the content are most similar in meaning to the query (semantic search).[38] The weights in the vectors are not adjusted, but are frozen embeddings. Subsequently, the most relevant parts of the content, those with the highest similarity scores, are retrieved and passed on.[39] Then, the information from the retrieved documents will be formatted and curated into a context that the LLM can use (enhanced context).[40] Finally, the LLM will generate an answer, based on the user's prompt.[41]

### *III. Generative Search*

Imagine a bustling library where each book represents a website and visitors come to browse, research, and take notes. Initially, people wander through the library's aisles, flipping through books to find information they need. One day, a librarian sets up a desk at the entrance with a sign: "Ask me anything." Visitors can pose their questions, and the librarian provides a list of books and their locations. Over time, the librarian becomes more advanced. Instead of merely pointing visitors to books, he summarizes the content directly at the desk. Visitors still receive a list of book references, but most are less inclined to visit the shelves themselves.

Ultimately, the librarian evolves into an omniscient entity. By leveraging knowledge gained from scanning every book in the library—and even external libraries—the librarian can answer any question instantly, without referencing specific books or shelves. Visitors stop entering the library altogether, as the librarian now fulfills all their needs at the entrance.

Similarly, generative search diminishes traffic to websites by providing users with answers directly, rendering the original sources less visible and visited.

A similar development one can observe, between the ranked list of URLs of a conventional search engine, the generative search results of Google, Baidu, and Perplexity AI, and generative results of a generative AI-search result. More positively, generative search has broken up the hegemony of Google as the search champion. This provides chances for other providers.[42]

A special application of RAG is generative search. Generative search results from Google, Perplexity AI, or Baidu do not merely display ranked links, but synthesize information across many sources and generate coherent, detailed responses, addressing complex queries in a conversational manner. "[T]raditional search engines that simply provide hyperlinks promote merely the discovery of copyrighted content, and not its substitution."[43]

---

[35]Don Woodlock, *What is RAG? (Retrieval Augmented Generation)*, (YouTube, Jan. 18, 2024), https://www.youtube.com/watch?v=u47GtXwePms; Harry Surden, *ChatGPT, AI Large Language Models, and Law*, 92 Fordham L. Rev. 1941, 1969 (2024) (augmenting a prompt is the idea of adding extra information before or after the prompt that GPT-4 can look at when it is deciding to predict its response).

[36]*What Is RAG (Retrieval-Augmented Generation)?*, AWS, https://aws.amazon.com/what-is/retrieval-augmented-generation/ (last visited Jan. 14, 2025).

[37]*Id.*

[38]*Id.*

[39]*Id.*

[40]*Id.*

[41]*Id.*

[42]Andrew Ellson, *Shoppers Ditch Search Engines as AI Shakes Up Online Spending*, The Times (Nov. 26, 2024), https://www.thetimes.com/uk/technology-uk/article/shoppers-ditch-search-engines-as-ai-shakes-up-online-spending-q6swp9hqw; *see also* Matteo Wong, *The AI Search War Has Begun: And Tech Companies Might Not Be the Winners*, The Atlantic (July 30, 2024), https://www.theatlantic.com/technology/archive/2024/07/perplexity-ai-search-media-partners/679294/.

[43]Dow Jones Complaint, *supra* note 34, at 5.

Both Google and Bing have integrated AI to allow users to engage in multi-step conversations with the search engine, refining and narrowing down their queries. This conversational search experience, powered by natural language understanding, is a key aspect of this generative search era.

Google's generative search results of "Google AI Overviews" refer to the integration of AI-powered language models, like Google's LaMDA or the integration of technologies similar to OpenAI's GPT-4, into its search engine. The results have been sometimes less than perfect.[44]

SearchGPT is a temporary prototype of new AI search features that aims to give rapid answers from relevant sources.[45] SearchGPT is designed to help users connect with publishers by prominently citing and linking to them in searches, with in-line, named attribution and links so users know where information is coming from and can quickly engage with even more results in a sidebar with source links. OpenAI is planning to integrate the best of these features into ChatGPT.

Initially, on May 10, 2023, users could start using Google Search Generation Experience (SGE) by activating it via Google Labs.[46] Individual Google Search users could activate SGE if they choose to do so.[47]

A year later, Google launched AI Overviews in the U.S.[48] Three months later, they announced that AI Overviews was rolled out to six more countries: Brazil, India, Indonesia, Japan, Mexico, and the UK.[49] On October 28, Google provided AI Overviews to over 100 countries.[50]

In August 2022, Perplexity AI started as a Discord bot connected to Bing; with 15 million active users, it is already valued 9 billion US dollars.[51] After allegations of plagiarism,[52] Perplexity AI hedged its bets and started to conclude licensing agreements with publishers, such as Time, Der Spiegel, and Fortune.[53] Nevertheless, Perplexity AI was sued by Dow Jones & Company, who described Perplexity's business model as "usurp[ing] content creators' monetization opportunities for itself."[54]

## C. Technological and Legal Measures Against RAG

This section explores the combined use of technological and legal measures to address the challenges posed by RAG systems. As RAG models increasingly rely on web content for training and output generation, website holders face significant risks, including copyright infringement, loss of revenue, and diminished traffic. This section outlines strategies that content providers and website holders can employ to protect their interests, emphasizing the need for targeted technical tools and enforceable legal frameworks. By integrating these measures, stakeholders can establish clear boundaries for permissible use while mitigating the exploitation of their content by RAG systems.

[44]Kyle Orland, *Google's "AI Overview" Can Give False, Misleading, and Dangerous Answers*, Ars Technica (May 24, 2024), https://arstechnica.com/information-technology/2024/05/googles-ai-overview-can-give-false-misleading-and-dangerous-answers/.

[45]*SearchGPT Prototype*, OpenAI, July 25, 2024, https://openai.com/index/searchgpt-prototype/ [hereinafter OpenAI].

[46]Elizabeth Reid, *Supercharging Search with Generative AI*, Google: The Keyword (Mar. 10, 2023), https://blog.google/products/search/generative-ai-search/. *See also* David Pierce, *Google Is Redesigning Its Search Engine—And It's AI All the Way Down*, The Verge (May 14, 2024), https://www.theverge.com/2024/5/14/24155321/google-search-ai-results-page-gemini-overview.

[47]Reid, *supra* note 46.

[48]Elizabeth Reid, *Generative AI in Search: Let Google Do the Searching for You*, Google: The Keyword (May 14, 2024), https://blog.google/products/search/generative-ai-google-search-may-2024/.

[49]Hema Budaraju, *New Ways to Connect to the Web with AI Overviews*, Google: The Keyword (Aug. 15, 2024), https://blog.google/products/search/new-ways-to-connect-to-the-web-with-ai-overviews/. *See also* Barry Schwartz, *Google AI Overviews Gains New Citations and Links, Plus More*, Search Engine Land (Aug. 15, 2024), https://searchengineland.com/google-ai-overviews-gains-new-citations-and-links-plus-more-445211.

[50]Barry Schwartz, *Google Expands AI Overviews to over 100 Countries*, Search Engine Land (Oct. 28, 2024), https://searchengineland.com/google-expands-ai-overviews-to-over-100-countries-447832.

[51]Alexandra Tremayne-Pengelly, *A.I. Answer Engine Maker Perplexity Is Valued at $9B After Latest Fundraising*, Observer (Dec. 19, 2024), https://observer.com/2024/12/perplexity-ai-valuation-9b/.

[52]Randall Lane, *Why Perplexity's Cynical Theft Represents Everything That Could Go Wrong With AI*, Forbes (June 11, 2024), https://www.forbes.com/sites/randalllane/2024/06/11/why-perplexitys-cynical-theft-represents-everything-that-could-go-wrong-with-ai/.

[53]Kylie Robison, *Perplexity is Cutting Checks to Publishers Following Plagiarism Accusations*, The Verge (July 30, 2024), https://www.theverge.com/2024/7/30/24208979/perplexity-publishers-program-ad-revenue-sharing-ai-time-fortune-der-spiegel.

[54]Dow Jones Complaint, *supra* note 34, at 5.

### *I. Technological Measures Opting Out*

Website holders who do not want their information to be used by RAG models often still want to be indexed by search engines, and therefore hesitate to block them. If they were to do so, their findability would decrease, leading to a significant drop in website traffic, which would adversely affect their advertising-derived income. Generative search is separate from training Large Language Models (LLMs). Therefore, websites can surface in search results even if they opted out of generative AI training.[55] Most websites have been dependent on Google Search for their traffic. Therefore, in any opt-out method, a distinction should be made between blocking spiders for training LLMs and being ranked in Google Search.

If a website holder wants her website to be indexed and ranked by search engines but not used by RAG models, she could apply a combination of technical and legal methods.

The Terms of Use can make the legal boundaries of content providers clear: for example, that scraping for LLM training or retrieving data for RAG models is not allowed. The problem, however, is that bots do not have the semantic understanding of the Terms of Use text yet. Even if they had this, then the implementation thereof depends on whether the bot is programmed ethically or maliciously.

The "robots.txt" file can guide search engine crawlers to the parts of the site that can be indexed. By disallowing certain sections or pages in the "robots.txt" file, one can prevent search engines from crawling or indexing certain content. However, RAG models may ignore these rules,[56] so this approach alone would not fully prevent them from accessing the website's data. In addition, one can use X-Robots-Tags, which are directives in Hypertext Transfer Protocol (HTTP), such as "noindex" or "nofollow," that allows one to control the indexing of non-Hypertext Markup Language (HTML) content such as PDFs, images, or other file formats.[57]

One can also prevent RAG systems from accessing or interacting with the website's Application Programming Interface (API) or data endpoints.[58] Other ways include limiting the number of requests per IP address, user, or token within a certain timeframe. This would prevent bots or automated systems like RAG from sending high-frequency requests to harvest data.[59] It also prevents Internet Protocol addresses (IP) from white-listing,[60] or black-listing,[61] blocking known IPs associated with unwanted bots or scraping activities. By requiring CAPTCHA ("Completely

[55]OpenAI, *supra* note 45.

[56]Elizabeth Lopatto, *Perplexity's Grand Theft AI*, The Verge (June 27, 2024), https://www.theverge.com/2024/6/27/24187405/perplexity-ai-twitter-lie-plagiarism (discussing Perplexity's use of scrapers that ignore the robots.txt protocol).

[57]Natalie Hoben, *Everything You Need To Know About The X-Robots-Tag HTTP Header*, Search Engine J. (Dec. 28, 2022), https://www.searchenginejournal.com/everything-you-need-to-know-about-the-x-robots-tag/314561/.

[58]Jamie Juviler, *API Security Best Practices: 10+ Tips to Keep Your Data Safe*, HubSpot (June 12, 2023), https://blog.hubspot.com/website/api-security. (noting that being part of an API's URL path is characteristic of endpoints: for example, in https://api.example.com/v1/products, "/v1/products" is the endpoint for retrieving information about products).

[59]*Web Application Firewall: Rate Limiting Rules, Best Practices*, Cloudflare Docs, https://developers.cloudflare.com/waf/rate-limiting-rules/best-practices (last visited Jan. 14, 2025).

[60]*IP Whitelisting: Basics, Alternatives and Beyond*, Sensfrx.ai (Dec. 12, 2024), https://blog.sensfrx.ai/ip-whitelisting/.

[61]*IP Blacklisting vs. IP Whitelisting, What's the Difference?*, This vs. That, https://thisvsthat.io/ip-blacklisting-vs-ip-whitelisting (last visited Jan. 14, 2025).

Automated Public Turing test to tell Computers and Humans Apart") or other challenge-response authentication systems, automated access to APIs can be prevented,[62] while API usage can monitor patterns through logging and analytics.[63]

Website holders can use encryption of sensitive data in transit with Hypertext Transfer Protocol Secure (HTTPS),[64] and tokenization, a process that replaces sensitive or important data with tokens that are meaningless on their own but can be mapped back to the original data only by an authorized system with the proper mechanism.[65]

Website holders can serve dynamic content that changes frequently or requires interaction, making it harder for RAG to retrieve and process the data.[66] They can also watermark and embed hidden metadata or other content tracking systems that can identify content when it is used in RAG models.[67]

RAG system providers can prevent liability for the deployment of infringement of copyrighted works by concluding licensing deals with content providers, and respecting the websites Terms of Use and technological protection measures.

Copyright holders should be able to send platforms that use a RAG system, which is infringing their copyrighted works, a cease-and-desist notice, specifying the material that is infringing and providing proof of ownership. AI service providers have a published process for handling intellectual property complaints, similar to a notice-and-takedown procedure of online service providers, under the Digital Millennium Copyright Act of 1998.[68] Outside of the U.S. similar systems are being used.[69] Service providers that use a RAG system, might file a counter-notice if they believe their use is fair use.[70]

### *II. Legal Claims Against Generative Search*

In October 2024, Dow Jones & Co. sued Perplexity AI, Inc. for the following claims:[71] copyright infringement,[72] for ingesting plaintiffs' copyrighted works to feed its RAG Index (count one); and to generate outputs for user queries (count two).

Count three refers to the claim of false designation of origin and dilution of plaintiffs' trademarks,[73] in case of hallucinated output after requests by users to provide for example the main news at the Wall Street Journal (owned by Dow Jones & Co.).

[62]Abhinav Bannerjee & Fenil Patel, *Use AWS WAF CAPTCHA to Protect Your Application Against Common Bot Traffic*, AWS (June 1, 2023), https://aws.amazon.com/blogs/security/use-aws-waf-captcha-to-protect-your-application-against-common-bot-traffic. *See also What Is reCAPTCHA?*, GOOGLE FOR DEVS., https://developers.google.com/recaptcha (last visited Jan. 14, 2025).

[63]*API Logs: Your First Line of Defense in Cybersecurity*, APITOOLKITS (Nov. 15, 2023), https://apitoolkit.io/blog/api-logs-in-cybersecurity/.

[64]Jessica Ofem, *What Is HTTPS?*, CLOUDFLARE, https://www.cloudflare.com/learning/ssl/what-is-https/ (last visited Jan. 14, 2025).

[65]Chris Brook, *What Is Data Tokenization? Key Concepts and Benefits*, DIGIT. GUARDIAN (Jan. 6, 2025), https://www.digitalguardian.com/blog/what-data-tokenization-key-concepts-and-benefits.

[66]Assad Abbas, *Keeping Data Safe: How to Counter Web Scraping Attacks*, TECHOPEDIA (Aug. 31, 2023), https://www.techopedia.com/keeping-data-safe-how-to-counter-web-scraping-attacks.

[67]Peizhuo Lv, Mengjie Sun, Hao Wang, Xiaofeng Wang, Shengzhi Zhang, Yuxuan Chen, Kai Chen & Limin Sun, *RAG-WM: An Efficient Black-Box Watermarking Approach for Retrieval-Augmented Generation of Large Language Models*, ARXIV 3 (Jan. 9, 2025), https://arxiv.org/pdf/2501.05249 (explaining how RAG-WM embeds structured watermark data (entity–relationship tuples) in a knowledge base that persists even after processing by external RAG systems. Because owners usually observe only outputs, the method relies on black-box detection: targeted queries reveal whether watermark information appears in the responses. The watermark is designed to survive paraphrasing, content removal, and knowledge expansion. If such markers appear in a system's answers, this suggests unauthorized use of the original knowledge base).

[68]17 U.S.C. § 512.

[69]Danny Friedmann, *Oscillating from Safe Harbor to Liability: China's IP Regulation and Omniscient Intermediaries*, *in* THE OXFORD HANDBOOK OF ONLINE INTERMEDIARY LIABILITY 277, 279–80, (Giancarlo Frosio ed., 2020).

[70]17 U.S.C. § 512(g).

[71]Dow Jones Complaint, *supra* note 34, at 34–38.

[72]17 U.S.C. § 106.

[73]15 U.S.C. § 1125 (including cases of trademark dilution by blurring and tarnishment).

Subsection 1 goes deeper into the claim of copyright infringement and looks at fair use. Subsection 2 deals with the non-legal norms against plagiarism and introduces "semi-plagiarism." Subsection 3 provides an additional claim, overseen by the plaintiffs, of unfair competition, where plagiarism and semi-plagiarism can play an auxiliary role.

Subsection 4 deals with the possible claims of unjust enrichment and negligence. The more generative search providers are being sued, the greater the interest for them to preempt these costly and time-consuming events; therefore Subsection 5 deals with licensing agreements. Finally, Subsection 6 provides a possible solution: agentic AIs, almost like a *deus ex machina*.

### 1. Copyright Infringement

To see whether there is copyright infringement, there should be substantial similarity and access.[74] Substantial similarity can be assessed at different abstraction levels (chapters, sections, paragraphs, sentences, and words, letters). Changing each word with a synonym, does not necessarily immunize one from copyright infringement. For example, if one is copying details of a plot, it could still be copyright infringement.[75] And even though ideas are non-protectable under copyright, if they are ever more refined, there will come a moment when these ideas become expressions.[76] In 1985, the Supreme Court in *Harper & Row v. Nation Enterprises* held that *The Nation*'s use of the paraphrased material still violated copyright because the essence of the copyrighted work—the "heart" of the expression—was taken.[77]

To paraphrase original human-created content does not necessarily indemnify a RAG from copyright infringement.[78]

Dow Jones & Co. argued because of the grand scale reproduction and/or derivative content paraphrased or not, "[t]he illegality of this massive copyright violation at the input stage does not depend on whether the particular outputs of Perplexity's so-called 'answer engine' are sufficiently similar in each instance to the copyrighted works of Plaintiffs as to constitute identical reproductions of those works."[79]

Generative search companies argue that fair use should be used as an affirmative defense against copyright infringement. Let us compare this with an analogous scenario similar to Google Overviews. In 1977, the plaintiff in *Wainwright Securities v Wall Street Transcript*, was providing reports on the stock market and the defendant used some opinions on certain sectors and stocks from the Wainwright report.[80] The question was whether the Wall Street Transcript's abstracts were fair use.[81] The fair use test consists of four elements that a court should consider;[82] the elements are not exhaustive and need not be applied mechanistically.[83]

---

[74]Skidmore v. Led Zeppelin, 952 F.3d 1051 (9th Cir. 2020) (revoking the inverse ratio rule between substantial similarity and access.).

[75]Richard Posner, The Little Book of Plagiarism 13 (2007).

[76]Danny Friedmann, *Creation and Generation Copyright Standards*, 14 N.Y.U. J. Intell. Prop. & Ent. L. 65 (2024).

[77]Harper & Row v. Nation Enters., 471 U.S. 539, 544, 565, 600 (1985).

[78]Salinger v. Random House, Inc., 818 F.2d 252, 253 (2d Cir. 1987) (making clear that a paraphrase can be infringing); *Wright v. Warner Books, Inc.*, 953 F.2d 731, 738 (2d Cir. 1991) (paraphrasing can convey the "heart" of a copyrighted work).

[79]Dow Jones Complaint, *supra* note 34, at 4.

[80]Wainwright Sec., Inc. v. Wall St. Transcript Corp., 558 F.2d 91, 94 (2d Cir. 1977).

[81]*Id.*

[82]17 U.S.C. § 107 (listing the elements:

> (1) the purpose and character of the use, including whether such use is of a commercial nature or is for non-profit educational purposes; (2) the nature of the copyrighted work; (3) the amount and substantiality of the portion used in relation to the copyrighted work as a whole; and (4) the effect of the use upon the potential market for or value of the copyrighted work.).

[83]*Harper & Row*, 471 U.S. at 588 ("These factors are not necessarily the exclusive determinants of the fair use inquiry and do not mechanistically resolve fair use issues.").

Judge Lasker of the Southern District of New York (SDNY) found that Wall Street Transcript's abstracts did not constitute fair use because: (1) the takings were "substantial in quality, and absolutely, if not relatively, substantial in quantity." So, the third fair use factor favors the plaintiff; (2) publication of the abstracts probably reduced the value of the research reports; this refers to the fourth fair use factor, which favors the plaintiff as well.[84] The District Court for the SDNY decision was affirmed by the Second Circuit. The argument made by Wall Street Transcript was that the reports consisted of financial news events: facts. One can argue that one needs to make a distinction between news events and how these events were expressed. Nevertheless, since *Salinger v. Colting* in 2010, the Second Circuit has decided to abrogate the case and held that transformativeness should play a bigger role than derivativeness in such a case.[85]

To see whether Google Overviews is fair use, let us also compare it with earlier Google services.

With Google News one can at least argue it is fair use if it only uses newspaper headings.[86] Because users will still have to click on the websites of the newspapers, they will be exposed to the advertisement—or in case of content behind a paywall, to be prompted to subscribe. The Google News case (AFP v. Google, Inc.) was settled between the parties.[87] In contrast, the Second Circuit decided that Google's digitization of books, including orphan works and out-of-print works and display of "snippets" in the Google Books project, was fair use.[88] It held them to be transformative: One cannot enjoy snippets of a book in the same way as reading an e-book. However, the generative search result can be enjoyed as the text on the website itself, and is arguably therefore not fair use.

### 2. Plagiarism and Semi-Plagiarism

Some generative search results do not provide sources. This lack of attribution leads to plagiarism,[89] which Posner describes as academic fraud.[90] Fraud is a tort, and often a crime, but plagiarism is neither.[91] In case of most plagiarism two parties are harmed: the original author of the content who was deprived of his right of attribution, and the readers that were deceived about the source of the content. Posner points out that this only matters when the readers actually care about the deception.[92]

Copyright only protects expressions of ideas, but plagiarism can also protect the right of the author to be recognized as the author of ideas, concepts or the reporter of facts.

Other generative search results, such as those of Google AI Overviews and Perplexity AI, do provide sources, but it is not always clear to which parts of the result they refer. This is what this author calls "semi-plagiarism." In the case of Google AI Overviews, sources are provided more as references that are placed at the end of a text than as footnotes that refer to particular segments within the text. In the case of Perplexity AI, the generated answer is presented followed by the sources, presented as tiny buttons, which have a clickable number corresponding to a particular segment. It seems clear that the generative search provider does not want to encourage the user to actually click on these links, otherwise it would have made the text itself clickable. Users make use of this constellation, seemingly oblivious or indifferent to the implications to the sourced websites.

[84]H. C. Wainwright & Co. v. Wall St. Transcript Corp., 418 F. Supp. 620, 625 (S.D.N.Y. 1976), *aff'd*, 558 F. 2d 91 (2d Cir. 1977), *cert. denied*, 434 U.S. 1014 (1978).

[85]Salinger v. Colting, 607 F. 3d 68, 73–74, 83 (2d Cir. 2010).

[86]Google's Motion and Memorandum for Partial Summary Judgment Dismissing Count II for Lack of Protectable Subject Matter at n.3, Agence France Press v. Google, Inc., No. 05-cv-00546 (D.C. Cir. Oct. 12, 2005) (Google arguing that using a headline to identify a news article was a classic fair use. The case settled before the court ruled on the issue).

[87]Stipulation of Dismissal at 1, *Agence France Press v. Google, Inc.*, No. 05-cv-00546 (D.C. Cir. Apr. 6, 2007).

[88]*Authors Guild v. Google*, 804 F.3d 202, 212 (2d Cir. 2015).

[89]*Plagiarize*, MERRIAM-WEBSTER DICTIONARY, https://www.merriam-webster.com/dictionary/plagiarize (last visited Aug. 26, 2025) ("[T]o steal and pass off (the ideas or words of another) as one's own: use (another's productions) without crediting the source.").

[90]POSNER, *supra* note 75, at 33.

[91]*Id.* at 34.

[92]*Id.* at 19–20.

To avoid copyright issues, in January 2024, Aravind Srinivas, CEO of Perplexity AI, has made his intention clear to attribute every part of a generated answer.[93] However, Randall Lane, Chief Content Officer and Editor of Forbes, wrote about two journalists from Forbes that had been reporting on a developing story on former Google CEO Eric Schmidt's secretive drone project. On June 6, they reported on ongoing testing of the drones in Silicon Valley. The very next day, Perplexity AI used its generated search to repackage the story and sent it to its subscribers via a mobile push notification. In addition, it created an AI-generated podcast based on the Forbes journalists' work, giving credits to Forbes only via small "F" buttons.[94]

### 3. Unfair Competition

*International News Service v. Associated Press* (1918) (*INS v AP*)[95] can shed light on the question whether generative search is fair competition. During WWI, the allied forces did not want to provide INS with news because its owner, William Randolph Hearst, favored Germany. Therefore, INS hired people to read East Coast newspapers that included AP news articles and subsequently to telegraph these to INS so that these parts could be published by its West Coast newspapers. Justice Oliver Wendell Holmes, dissenting, asserted that there was no property in news, because the news is merely facts,[96] and that there was no passing off (misrepresentation to deceive readers), as would be necessary to reach a finding of unfair competition.[97] Even though the INS might have given attribution to AP, the Supreme Court held that there was still misappropriation of "hot news," and qualified it as unfair competition.[98] The court held that the investments in news-gathering need to be protected. This is a sweat of the brow argument, probably inspired by the realization that in news-gathering a certain degree of sharing among journalists should be allowed, but not to an excessive degree. Too much free-riding would make investigative journalism, which is labor intensive indeed, no longer worthwhile.[99] Therefore, the plagiarism or semi-plagiarism concomitant with generative search can be qualified as deceptive and *a fortiori* as a misrepresentation.

### 4. Unjust Enrichment and Tort of Negligence

Google Search has been the dominant intermediary between internet users and websites, almost worldwide.[100] For a decade, over 90 percent of internet users made use of Google Search.[101] Website holders, too, were dependent on Google Search for the vast majority of their traffic referrals.[102]

[93]Office of Alumni and Corporate Relations IIT Madras, *IITM ACR - LLS - TALK BY SHRI. ARAVIND SRINIVAS – FOUNDING STORY AND JOURNEY OF PERPLEXITY*, at 49:05 (YouTube, Jan. 18, 2024), https://www.youtube.com/watch?v=ygRVDIwheB4 ("The way we are thinking about it is we are at least attributing every part of the answer. Where are we getting it from in terms of inline citations as well as the source panel at the top.").

[94]Randall Lane, *Why Perplexity's Cynical Theft Represents Everything That Could Go Wrong With AI*, Forbes (June 11, 2024), https://www.forbes.com/sites/randalllane/2024/06/11/why-perplexitys-cynical-theft-represents-everything-that-could-go-wrong-with-ai/ (noting that Perplexity's podcast outranked Forbes' content on this topic within Google search).

[95]Int'l News Serv. v. Associated Press, 248 U.S. 215, 215 (1918).

[96]*Id.* at 246.

[97]*Id.* at 246–47.

[98]*Id.* at 235 ("The peculiar value of news is in the spreading of it while it is fresh; and it is evident that a valuable property interest in the news, as news, cannot be maintained by keeping it secret.").

[99]Shyamkrishna Balganesh, *Hot News: The Enduring Myth of Property in News*, 111 Colum. L. Rev. 419, 496 (2011).

[100]Except for China, Russia, and South Korea, where respectively Baidu, Yandex, and Naver are dominant. Kaan Güner, *3 Countries Where Google Isn't the Top Search Engine*, PolarMass (Sep. 3, 2024), https://polarmass.com/blog/3-countries-where-google-isnt-the-top-search-engine/.

[101]Danny Goodwin, *Google's Search Market Share Drops Below 90% for First Time Since 2015*, Search Engine Land (Jan. 13, 2025), https://searchengineland.com/google-search-market-share-drops-2024-450497.

[102]*See* Rand Fishkin, *Who Sends Traffic on the Web and How Much? New Research from Datos & SparkToro*, SparkToro (Mar. 11, 2024), https://sparktoro.com/blog/who-sends-traffic-on-the-web-and-how-much-new-research-from-datos-sparktoro ("63.41 percent of all U.S. web traffic referrals from the top 170 sites initiated on Google.com").

Millions of websites holders have placed Google Display Ads on their websites to earn revenues.[103] These website holders are thus not only dependent on Google for traffic referrals but are also in contractual relation with Google.

Google gradually implemented generative search: on May 10, 2023, Google tested Google Search Generation Experience (SGE) via Google Labs. Individual Google Search users could activate SGE if they choose to do so.[104] A year later, Google launched SGE's successor called AI Overviews in the U.S.[105] Three months later, Google announced that AI Overviews was rolled out to six more countries: Brazil, India, Indonesia, Japan, Mexico, and the UK.[106] On October 28, Google provided AI Overviews to over 100 countries.[107]

Google's implementation of generative search foreseeably harmed dependent companies by reducing internet users' visits to their websites. Because "Google is doing the searching for you,"[108] users are less likely to click through to external sites, undermining website holders' revenue streams, particularly those reliant on advertisements (mainly through Google) or paywalls.[109] This situation may give rise to legal claims based on restitution, unjust enrichment, and the tort of negligence, particularly in the context of a contractual relationship. Under the principle of unjust enrichment, "a person who has been unjustly enriched at the expense of another is required to make restitution."[110] It can be argued that a search engine provider, such as Google, which implements generative search and sends web crawlers to index websites for its RAG systems instead of directing traffic to those sites, should have anticipated a substantial decline in traffic. Given Google's dominant market share of approximately 90 percent, one could argue that it failed to exercise reasonable care in informing websites about the potential impact of these changes, thereby depriving them of crucial information and causing foreseeable harm.

### *5. Licensing Deals*

Generative search deprives website holders of the revenue that can be generated from internet users visiting their websites: the fee for subscriptions to have access to their website, or the pay per clicks on advertisements on their website. However, generative search providers can compensate website holders for this deprivation and indemnify them against liability for copyright infringement. A current leader of AI services, OpenAI, has stated that it is committed to "a thriving ecosystem of publishers and creators."[111] Since being sued by authors[112] and publishers in 2023,[113] OpenAI has started to conclude licensing agreements with a slew of publishers.[114] The Copyright Clearance Center announced to include some AI rights in its Annual Copyright License

[103] *Reach a Larger or New Audience with Google Display Network Targeting*, GOOGLE ADS (Mar. 20, 2023), https://ads.google.com/intl/en_us/home/resources/articles/reach-larger-new-audiences/.

[104] Reid, *supra* note 46.

[105] Reid, *supra* note 48.

[106] Buduraju, *supra* note 49.

[107] Schwartz, *supra* note 50.

[108] Reid, *supra* note 48.

[109] Henk van Ess, *How AI Bots Quietly Dismantle Paywalls via Web Search*, DIGITAL DIGGING (July 11, 2025), https://www.digitaldigging.org/p/how-ai-bots-quietly-dismantle-paywalls (circumventing paywalls using different techniques, including "conducting live searches (mining social media etc.) to actively reconstruct paywalled articles").

[110] *See generally* RESTATEMENT (FIRST) OF RESTITUTION (A.L.I. 1937).

[111] *SearchGPT Prototype*, OPENAI (July 25, 2024), https://openai.com/index/searchgpt-prototype/.

[112] *See generally* Class Action Complaint, Tremblay v. OpenAI, Inc., No. 23-cv-03223 (N.D. Cal. June 28, 2023); Complaint, Authors Guild v. OpenAI Inc., No. 23-cv-08292 (S.D.N.Y. Sep. 19, 2023).

[113] *See generally* Complaint, The New York Times Co. v. Microsoft Corp., No. 23-cv-11195 (S.D.N.Y. Dec. 27, 2023).

[114] Bill Rosenblatt, *The Media Industry's Race To License Content For AI*, FORBES (July 18, 2024), https://www.forbes.com/sites/billrosenblatt/2024/07/18/the-media-industrys-race-to-license-content-for-ai/ (identiftying Vox Media (parent of New York Magazine, The Verge, and Eater), News Corp (Wall Street Journal, New York Post, The Times (London)), Dotdash Meredith (People, Entertainment Weekly, InStyle), Time, The Atlantic, Financial Times, and European giants such as Le Monde of France, Axel Springer of Germany, and Prisa Media of Spain). *See also* Alexandra Bruell, Sam Schechner & Deepa

for corporations.[115] Large publishing houses can negotiate remuneration in exchange for their information, leveraging their scale and resources to secure favorable agreements. In contrast, small- and medium-sized enterprises (SMEs) that operate websites often lack the bargaining power and infrastructure to achieve similar outcomes. This disparity puts SMEs at a significant disadvantage, as they struggle to monetize their content effectively or protect it from unauthorized use by generative search systems. Consequently, this dynamic accelerates market concentration, consolidating power and revenue among dominant players while further marginalizing smaller content creators. Such an imbalance not only stifles competition but also undermines the diversity and innovation that SMEs bring to the digital ecosystem. To address this, regulatory interventions may be necessary to ensure equitable opportunities for all content providers.

### 6. Agentic AI

Fenwick, Jurcys, and Loikkanen posited AI agents as a possible remedy against the traffic and revenue loss of websites:[116] Google's AI agent could gather information from websites to generate search results and remunerate AI agents of these websites for this information.

While RAG is reactive, responding to input prompts with contextually relevant information retrieved from external sources, agentic AI is capable of autonomous decision-making and acting towards a goal, and typically can learn from experience.[117] RAG focuses on factual grounding, AI Agents provide planning capabilities and adaptability within complex environments.[118] Both systems can be merged into Agentic RAG, which combines RAG's knowledge capabilities with AI agents' decision-making skills.[119] Until this agentic-centric solution is implemented, generative search providers might expect lawsuits from authors, publishers and website holders.

## D. Summary Generative Search

The internet's shift from a decentralized, content-rich ecosystem to centralized generative platforms marks a pivotal change. While RAG-based search enhances user experience, it undermines website sustainability by reducing traffic. Website owners can adopt technological and legal strategies, but ongoing disputes, like *Dow Jones & Co. v. Perplexity AI*,[120] highlight unresolved issues such as copyright infringement and unfair competition. Licensing agreements and agentic AI offer potential solutions, though they remain underdeveloped.

Generative search contributes to the decline in website traffic, just as shadow banning can reduce visibility of individuals' posts on social media. Both the impact of generative search on traffic and the rules governing shadow banning remain opaque and lack transparency.

"It exists" is an eternalist view; "It does not exist" is an annihilationist idea. Therefore the wise one should not have recourse to either existence or nonexistence (Madhyamaka).[121]

---

Seetharaman, *OpenAI, WSJ Owner News Corp Strike Content Deal Valued at Over $250 Million*, Wall St. J. (May 22, 2024), https://www.wsj.com/business/media/openai-news-corp-strike-deal-23f186ba.

[115]*CCC Pioneers Collective Licensing Solution for Content Usage in Internal AI Systems*, Copyright Clearance CTR. (July 16, 2024), https://www.copyright.com/media-press-releases/ccc-pioneers-collective-licensing-solution-for-content-usage-in-internal-ai-systems/.

[116]Mark Fenwick, Paul Jurcys & Valto Loikkanen, *Welcome to Google Zero! Incentives, Remuneration & Copyright in a New World of AI Agents* 7 (Working Paper, 2024), https://papers.ssrn.com/sol3/papers.cfm?abstract_id=4844108.

[117]Adrien Payong & Shaoni Mukherjee, *RAG, AI Agents, and Agentic RAG: An In-Depth Review and Comparative Analysis*, Digit. Ocean (Jan. 14, 2025), https://www.digitalocean.com/community/conceptual-articles/rag-ai-agents-agentic-rag-comparative-analysis.

[118]*Id.*

[119]Aditi Singh, Abul Ehtesham, Saket Kumar & Tala Talaei Khoei, *Agentic Retrieval-Augmented Generation: A Survey On Agentic RAG*, arXiv 10 (Jan. 15, 2025), https://arxiv.org/pdf/2501.09136v1.

[120]Dow Jones Complaint, *supra* note 34.

[121]Nagarjuna's Middle Way: Mulamadhyamakakarika 153–163 (Mark Siderits & Shoryu Katsura trans., Simon & Schuster 2013) (150 C.E.).

Perhaps not the wise, but opportunistic platforms use shadow banning, operating in the algorithmic twilight zone between presenting search results and blocking them. This is how they can avoid the risk of being sued, and can optimize the copyright holders' interests to directly or in a time-phased way de-rank or reduce traffic to unauthorized but possibly legal content into relative oblivion. This midway manipulation of traffic to suspected content is opaque to the uploader of content, and the general user, lacks any redress mechanism, and possibly chills the freedom to share transformed content that includes copyrighted works. Oftentimes, the platform has license conditions which would immunize it for de-ranking or reducing traffic, without the need to provide reasons, which is arguably incompatible with fundamental rights.

## E. Shining a Light on Shadow Banning

To avoid liability for copyright infringement, platforms tend to err on the side of over-enforcement, often suppressing lawful but unauthorized user-generated content. This exposes them to potential legal claims from users whose expression is curtailed. Platforms increasingly deploy algorithmic enforcement mechanisms that operate with minimal transparency. This section examines how such systems implement stealth measures—including algorithmic dissuasion, de-ranking, and copyright strikes—to moderate content in ways that often escape public scrutiny and procedural challenge. It examines the legal and ethical challenges of these methods, particularly in balancing copyright enforcement with user rights, and fair use or exceptions and limitations to copyright. Focusing on China's Regulation on Algorithmic Recommendations (RAR) and the EU's Artificial Intelligence Act (AIA), this section highlights the impact of algorithm-driven moderation on content visibility, innovation, and free expression.

### *I. Algorithmic Dissuasion*

How to make algorithmic dissuasion transparent and assailable with regard to uploaded works that allegedly infringe copyright, but might be fair use or fall within an exception or limitation? This section uses China's Provisions on the Management of Algorithmic Recommendations in Internet Information Services, hereinafter Regulation on Algorithmic Recommendations (RAR),[122] as a case study. RAR is arguably the most far-reaching attempt to regulate aspects of recommendation algorithms.

#### *1. Reining in Recommendation Algorithms*

China's copyright system—which is a hybrid of exceptions and limitations, the three-step test, and fair use principles—provides an example of a system within which unauthorized works uploaded on a platform are not always copyright infringements. China's third amendment to the Copyright Law (2020) provides 13 limitations,[123] each subject to the three-step test of Article 24(1). The three-step test, based on Article 13 Agreement on Trade-Related Aspects of Intellectual Property Rights (TRIPs), which is an extension of Article 9.2 Berne Convention for the Protection of Literary and Artistic Works,[124] includes three open norms: "Members shall confine limitations or exceptions to exclusive rights to [1] certain special cases [2] which do not conflict with a normal exploitation of the work and [3] do not unreasonably prejudice the legitimate interests of the right holder."[125]

[122]Hulianwang Xinxi Fuwu Suanfa Tuijian Guanli Guiding (互联网信息服务算法推荐管理规定) [Provisions on the Management of Algorithmic Recommendations in Internet Information Services] [hereinafter RAR] (promulgated by the Cyberspace Admin. of China, Dec. 31, 2021, effective Mar. 1, 2022), https://www.lawinfochina.com/display.aspx?id=37324&lib=law.

[123]Zhonghua Renmin Gongheguo Zhuzuoquan Fa (2020 Xiuzheng) (中华人民共和国著作权法(2020修正)) [Copyright Law of the People's Republic of China (2020 Amendment)] (promulgated by the Standing Comm. Nat'l People's Cong., Nov. 11, 2020, effective June 1, 2021), https://www.lawinfochina.com/display.aspx?lib=law&id=9280&CGid=.

[124]Berne Convention for the Protection of Literary and Artistic Works, Sep. 9, 1886, 828 U.N.T.S. 221 (including the three-step test to the Berne Convention since the Stockholm Act of 1967, it was limited to the reproduction right of a work).

Article 21(1) number 13 of the Copyright Law of the PRC states: "Other circumstances provided for by laws and administrative regulations." However, since 2004 at least some Chinese courts[126] have been willing to apply fair use principles,[127] despite the lack of fair use in the Copyright Law of 2010[128] (and 2020). In 2011, the Supreme People's Court issued an opinion[129] where in the "special case" of uses that promote innovation and business development, a more flexible limitations regime is admissible, if it comports with four fair-use factors,[130] inspired by Section 107 of the US Copyright Law.[131]

Instead of blocking or deleting content that is not straightforward, counterfeit, or pirated, a platform can instruct its algorithm to allow uploading content but label it as suspect. Subsequently, the algorithm of the platform will put the search or feed results directly or after a certain period or in certain geographical areas as nearly unfindable by squeezing the traffic to the content.

### *2. Unauthorized but Legal User-Generated Content*

This article is investigating the legitimacy and chilling effect of this practice in light of RAR[132] and AIA,[133] and takes user-generated content that was created unauthorizedly but possibly legal if it did not infringe copyright as a case study. The results of this case study are to a large extent also applicable to content that includes one or more unauthorized but possibly legal trademark—but that is outside the scope of this Article.

[125]Agreement on Trade-Related Aspects of Intellectual Property Rights art. 13, Apr. 15, 1994, Marrakesh Agreement Establishing the World Trade Organization, Annex 1C, 1869 U.N.T.S. 299. (extending limitation to all exclusive rights that are protected under copyright). *See, e.g.*, Zhonghua Renmin Gongheguo Zhuzuoquan Fa (2010 Xiuzheng) (中华人民共和国著作权法(2010修正)) [Copyright Law of the People's Republic of China (2010 Amendment)] (promulgated by the Standing Comm. Nat'l People's Cong., Feb. 26, 2010, effective Apr. 4, 2010), https://english.www.gov.cn/archive/laws_regulations/2014/08/23/content_281474982987430.htm; 17 U.S.C. § 106. *See also* Directive 2019/790, of the European Parliament and of the Council of 17 April 2019 on Copyright and Related Rights in the Digital Single Market and Amending Directives 96/9/EC and 2001/29/EC, 2019 O.J. (L 130) 92–125 (providing exceptions and limitations that are subject to the three-step test).

[126]Seagull Haiyan Song, *Reevaluating Fair Use in China — A Comparative Copyright Analysis of Chinese Fair Use Legislation, the U.S. Fair Use Doctrine, and the European Fair Dealing Model*, 51 IDEA 453 (2011).

[127]Peter Ganea, Danny Friedmann, Jyh-An Lee & Douglas Clark, Intellectual Property in China 359–62 (Christopher Heath ed., 2d ed. 2021).

[128]Zhonghua Renmin Gongheguo Zhuzuoquan Fa (2010 Xiuzheng) (中华人民共和国著作权法(2010修正)) [Copyright Law of the People's Republic of China (2010 Amendment)] (promulgated by the Standing Comm. Nat'l People's Cong., Feb. 26, 2010, effective Apr. 4, 2010), https://english.www.gov.cn/archive/laws_regulations/2014/08/23/content_281474982987430.htm.

[129]Zuigao Renmin Fayuan Yinfa <Guanyu Chongfen Fahui Zhishi Chanquan Shenpan Zhineng Zuoyong Tuidong Shehui Zhuyi Wenhua Da Fazhan Da Fanrong He Cujin Jingji Zizhu Xietiao Fazhan Ruogan Wenti De Yijian> De Tongzhi (Fa Fa [2011] 18 Hao) (最高人民法院印发《关于充分发挥知识产权审判职能作用推动社会主义文化大发展大繁荣和促进经济自主协调发展若干问题的意见》的通知 ( 法发〔2011〕18号)) [Notice of the Supreme People's Court on Issuing the Opinions on Issues concerning Maximizing the Role of Intellectual Property Right Trials in Boosting the Great Development and Great Prosperity of Socialist Culture and Promoting the Independent and Coordinated Development of Economy (No. 18 [2011] of the Supreme People's Court)] (promulgated by the Judicial Comm. Sup. People's Ct., Dec. 16, 2011, effective Dec. 16, 2011) Sup. People's Ct. Gaz., Dec. 16, 2011, https://gongbao.court.gov.cn/Details/90857967e518c766c368851b1b705a.html.

[130]Ganea et al., *supra* note 127.

[131]17 U.S.C. § 107.

[132]RAR, *supra* note 122.

[133]Commission Regulation 2024/1689 of June 13, 2024, Laying Down Harmonised Rules on Artificial Intelligence and Amending Regulations (EC) No 300/2008, (EU) No 167/2013, (EU) No 168/2013, (EU) 2018/858, (EU) 2018/1139 and (EU) 2019/2144 and Directives 2014/90/EU, (EU) 2016/797 and (EU) 2020/1828 (Artificial Intelligence Act), 2024 O.J. [hereinafter AIA], https://data.europa.eu/eli/reg/2024/1689/oj.

Article 2 RAR provides examples of recommendation algorithm technologies that can generate and synthesize individualized pushing of content; refine the sequence of content (feed) and search filtering.[134]

The recommendation algorithms of platforms such as Google, YouTube, TikTok/Douyin, et cetera, extract a fraction of content and rank it, based on the interactions of the user on that platform and possibly on several other sites as well. The algorithms learn to read the content preferences and usage patterns, combined with the location and time of users. The resolution of the user profiles is becoming ever sharper. Paraphrasing Yanis Varoufakis, we are training the algorithm so that the algorithm can train us.[135]

Floridi has pointed out five ethical principles for algorithms; "beneficence, non-maleficence, autonomy, justice, and explicability."[136] Neuwirth demonstrates that these general ethical standards are not met with concrete safeguards against different forms of manipulation, for instance regarding subliminal AI systems.[137]

In the same vein, this Article asserts that there are neither concrete safeguards yet for exceptions and limitations in regard to the automatic content recognition algorithms, nor for de-ranking or the reduction of traffic.

Next to the twilight zone, as mentioned above caused by uncertain algorithmic moderation, one can argue that there is a second twilight zone: the automatic content recognition algorithm. If it labels content as unauthorized, it does not know whether it is legal and thus non-infringing. Automatic content recognition can filter copyright infringements at the moment of uploading or thereafter. These filtering algorithms, such as YouTube Content ID, are supposed to account for copyright exceptions and limitations in civil law jurisdictions and for fair use in common law countries—also recognized to some extent in China—but in practice they often fail to do so.[138] In the case of copyrighted content, the automatic content recognition algorithms of platforms in at least the U.S. and EU have transformed from under-enforcers with false negatives to over-enforcers with false positives,[139] to steer clear of the chance of being sued by copyright holders. In the case of false negatives, online service providers were unable or unwilling to take down unauthorized copyrighted works. The *Viacom Intern., Inc. v. YouTube, Inc.* case, heard by the U.S. Second Circuit Court of Appeals, arguably demonstrates willful blindness.[140] The founders of YouTube were taking their chances, after a notice-and-takedown request by the copyright holders, by failing to remove uploaded, copyright-infringing works which were increasing traffic to the site.[141]

Since the advent of technologies that more accurately filter unauthorized copyrighted works, both overenforcement and underperformance have occurred depending on the size of the online service provider. Bar-Ziv and Elkin-Koren point out that copyright holders normally focus their enforcement strategies on global intermediaries instead of local platforms.[142] This can lead to overenforcement, for the former, and underenforcement, for the latter, in the same time period. In the case of false positives, the online service providers were unable or unwilling after a notice-and-takedown request to make sure that the works were indeed infringing copyright by assessing whether the works were in the public domain or subject to limitations of fair use. The Ninth Circuit case *Lenz v. Universal Music Corp.* exemplifies this overprotection, where YouTube took down a homemade video of toddlers running around a kitchen because Prince's "Let's Go Crazy" was barely perceptible in the background.[143]

[134]RAR, *supra* note 122.

[135]CAMBRIDGE UNION, *Yanis Varoufakis | Cambridge Union*, at 46:47 (YouTube, Mar. 2, 2023), https://www.youtube.com/watch?v=6u4YdHAgwd4&t=17s ("Train it to train me to buy the things that you stock for me . . . .).

[136]Luciano Floridi & Josh Cowls, *A Unified Framework of Five Principles of AI in Society*, 1 HARV. DATA SCI. REV. at 4 (2019), https://hdsr.mitpress.mit.edu/pub/l0jsh9d1/release/8.

[137]ROSTAM NEUWIRTH, THE EU ARTIFICIAL INTELLIGENCE ACT: REGULATING SUBLIMINAL AI SYSTEMS 2 (2022).

[138]Danny Friedmann, *Digital Single Market, First Stop to The Metaverse: Counterlife of Copyright Protection Wanted*, *in* LAW AND ECONOMICS OF THE DIGITAL TRANSFORMATION 137, 141 (Klaus Mathis & Avishalom Tor eds., 2023).

[139]*Id.*

[140]Viacom Int'l, Inc. v. YouTube, Inc., 676 F.3d 19 (2d Cir. 2012).

[141]*Id.*

[142]Sharon Bar-Ziv & Niva Elkin-Koren, *Behind the Scenes of Online Copyright Enforcement: Empirical Evidence on Notice & Takedown*, 50 CONN. L. REV. 339, 344 (2018).

In addition to the deployment of filter algorithms to enforce against copyright infringements, search engines and platforms started to use algorithms to provide users with personalized recommendations. In 2003, Amazon made clear it used a recommendation algorithm for its e-commerce site.[144] The recommendation algorithm, owned by China's ByteDance Ltd., has garnered significant attention not only for potential national security risks and its susceptibility to foreign adversaries,[145] but also for its extraordinary effectiveness, which has been criticized for fostering user addiction.[146]

### II. De-Ranking or Reduction of Traffic

Most internet users will focus on the first page of search results, and few users dive any deeper. With feeds of video or music-sharing sites this often happens as well, although many users scroll a bit more. If one scrolls down, new feed results will load, again and again. Just as each search result, every feed will be fed by a recommendation algorithm, based on the user's historic profile that is updated with every click. The direct visibility of a search result using Google or Baidu is imperative for the chances that a user will find it. *A fortiori*, at least since 2012, Alphabet's Google has been penalizing sites that infringe copyright by delisting them from their index or downranking them and making them practically unfindable,[147] a practice that continued in 2023[148] and 2024.[149] The recommendation algorithms of Alphabet's YouTube and X Corp's X work in a similar way. These intentional reductions in visibility are comparable to shadow banning in content moderation.[150] The algorithm that dictates the degree of visibility of content on YouTube is determined by factors such as the click-through rates and average view duration, but also by whether works are unauthorized and whether there are copyright claims or strikes against the uploader.

[143] *See generally* Lenz v. Universal Music Corp., 801 F.3d 1126 (9th Cir. 2015).

[144] Greg Linden, Brent Smith & Jeremy York, *Amazon.com Recommendations: Item-to-Item Collaborative Filtering*, 7 IEEE Internet Computing 76 (2003). *See also* Larry Hardesty, *The History of Amazon's Recommendation Algorithm, Collaborative Filtering and Beyond*, Amazon Sci. (Nov. 22, 2019), https://www.amazon.science/the-history-of-amazons-recommendation-algorithm.

[145] TikTok, Inc. v. Garland, 604 U.S. 56 (2025) (citing the importance of national security and the platform's susceptibility to foreign adversary influence and data collection while upholding the Protecting Americans from Foreign Adversary Controlled Applications Act (which prohibits TikTok's U.S. operations unless divested from Chinese control)).

[146] Bobby Allyn, Sylvia Goodman & Dara Kerr, *Inside the TikTok Documents: Stripping Teens and Boosting 'Attractive' People*, NPR (Oct. 16, 2024), https://www.npr.org/2024/10/12/g-s1-28040/teens-tiktok-addiction-lawsuit-investigation-documents (noting that watching the app becomes a habit in under thirty-five minutes, or 260 videos). Shaun Shia, *Recommendation System Behind Toutiao News App*, Medium (Jul. 15, 2018), https://medium.com/@impxia/newsfeed-system-behind-toutiao-2c2454a6d23d (ByteDance started with recommendation algorithms in 2012, when it launched news aggregator Toutiao).

[147] Danny Sullivan, *The Pirate Update: Google Will Penalize Sites Repeatedly Accused Of Copyright Infringement*, Search Engine Land (Aug. 10, 2010), https://searchengineland.com/dmca-requests-now-used-in-googles-ranking-algorithm-130118.

[148] Tatiana Tsyulia, *15 Major Google Penalties and How to Fix Them*, SEO PowerSuite (Feb. 13, 2023), https://www.link-assistant.com/news/google-penalties-guide.html.

[149] *In What Ways Can Copyright Infringement Affect SEO Success in 2024?*, JEMSU, https://jemsu.com/in-what-ways-can-copyright-infringement-affect-seo-success-in-2024/ (last visited Jan. 26, 2025).

[150] Paddy Leerssen, *An End to Shadow Banning? Transparency Rights in the Digital Services Act Between Content Moderation and Curation*, 48 Comput. L. & Sec. Rev., May 2023, at 1, 2 (using the term "shadow banning" for content moderation sanctions which are undetectable to those affected, and dealing with the EU Digital Services Act's due process and transparency in regard to content moderation).

### *III. Copyright Claim and Copyright Strike*

YouTube Content ID lets the copyright holder claim the content. Instead of submitting a notice and take-down request, he or she can track the video's performance, block it on a country-by-country basis, or choose to monetize the video by placing ads on it. During the "tracking performance" phase, "[t]he claim may affect [the views of the uploader's content] as the [copyright] owner may restrict the video from appearing on certain websites, devices, or even various countries."[151]

> YouTube does something like a soft ban. Your video's [sic] lose rankings on search terms or sometimes even do not show up on search results. There is no proven way to sort this out. You could try changing the title and headlines of the videos and see if it starts appearing on search results again or just wait it out and see if the views increase after two to three weeks.[152]

A copyright strike is when YouTube receives a complete and valid legal takedown request from the copyright holder,[153] then takes down the video to comply with copyright law. After three strikes, the account and any associated channels are subject to termination, and all videos that were uploaded will be removed.

After the first strike, the uploader needs to go through an online copyright course. He or she can resolve the copyright strike by waiting for it to expire after 90 days, contacting the owner and convincing him or her to withdraw the takedown request, or submitting a counternotification opposing the claim that there is copyright infringement. Regarding the lattermost, for example, the uploader can claim that the work is in the public domain,[154] that the uploader is the copyright holder of the work, that the copyright holder has granted permission to use the work, that an exception or limitation applies, or that the use can be qualified as fair use/fair dealing.

After the first and second strike, YouTube may reduce the visibility of the channel and its videos in search results and reduce recommendations, which means fewer views and subscribers. If the uploader receives three strikes, his or her account is subject to termination to prevent repeat copyright infringement.

## F. Algorithmic Dissuasion: De-Ranking Possible Copyright-Infringing Content into Oblivion

This section explores the interactions between algorithmic dissuasion and IP regulation across three distinct contexts: the interests of stakeholders, China's regulatory framework, and the EU's approach to algorithmic governance. Subsection I examines the key actors—copyright holders, platforms, and users—who are affected by algorithmic dissuasion, highlighting the strategic motivations and challenges surrounding the suppression or promotion of suspect content. Subsection II focuses on China's RAR, emphasizing its balance of transparency, user rights, and enforcement while addressing IP concerns. Subsection III evaluates the EU's AIA, assessing its risk-based regulatory framework and its implications for algorithmic content recognition in relation to fundamental rights and copyright protections. Together, these sections provide a comprehensive analysis of the legal and practical tensions arising from algorithmic decision-making in the context of IP.

[151]Mary Woodcock, *Can I Still Monetize With a Copyright Claim?*, LICKD (Aug. 16, 2022), https://lickd.co/blog/music-licensing/can-i-still-monetize-with-a-copyright-claim.

[152]Mark Dsouza, *After a copyright strike, views are decreasing from my channel. Does Copyright Affect YouTube Channels?*, QUORA (2021), https://qr.ae/prxxk3 (last visited Aug. 27, 2025).

[153]*Copyright Strike Basics*, GOOGLE, https://support.google.com/youtube/answer/2814000?hl=en (last visited Aug. 27, 2025).

[154]Timothy Geigner, *White Noise On YouTube Gets FIVE Separate Copyright Claims From Other White Noise Providers*, TECHDIRT (Jan. 5, 2018), https://www.techdirt.com/2018/01/05/white-noise-youtube-gets-five-separate-copyright-claims-other-white-noise-providers/.

### *I. Instigators, Perpetrators, and Victims of Algorithmic Dissuasion*

To cast light on the method of fading out suspect content, this section will look at the effects on the respective protagonists of algorithmic dissuasion: copyright holders, platforms, and users.

There can be different reasons for the interest of copyright holders in reducing but not completely blocking unauthorized content, whether illegal or legal. In the interest of promotion of, for example, a movie, its copyright holder might prefer to allow at least for a certain period that a movie, or parts of that movie, can be shared by YouTube channels that cater to early adopters, whether these videos are infringing copyright or not. Platforms such as YouTube can recommend such videos to those YouTube users that meet for example the profile of early adopters. However, once the movie debuts in cinemas in a certain region, the copyright holder might want the traffic reduced to all unauthorized videos, including those that are legal.

Another reason is that copyright holders might not see it as conducive to alienate users that are also potential consumers by blocking their content, which could backfire on social media and become a public relations debacle.

A platform might be inclined to reduce the traffic to content that uses unauthorizedly copyrighted works. The platform's automatic content recognition algorithm cannot yet apply the three-step test and concepts such as parody and transformativeness, so it cannot detect unauthorized but legal content. Platforms have only a nominal safe harbor and are still being sued by copyright holders. Even if the copyright holders can find the direct infringer and he or she is in the same jurisdiction, he or she might not have the financial means to pursue a lawsuit, while the platform has deep pockets.[155] By reducing traffic to unauthorized content, platforms reduce the chance of being sued.

With impunity, opportunistic platforms seem to be able to reduce the traffic to users that unauthorizedly uploaded a copyrighted work, no matter whether this was legal or not. Because of the opaqueness of this process, users can hardly detect that this is happening. And if they suspect it, they cannot prove it. Information about such events is also known as dark data.[156] A decrease in the number of visitors to a certain uploaded content on a platform can be due to several reasons related to the relevance ranking algorithm of the platform—which has a number of components, all trade secrets. There is obviously no redress mechanism for something that cannot be detected with certainty. In addition, there is a contractual relationship between the platform and the uploader, where the first limits its liability and the latter *de facto* agrees with any curatorial interventions,[157] but which arguably is not compatible with fundamental rights, such as the right of freedom of expression.

> Limitation of liability
>
> 1. ERRORS, MISTAKES, OR INACCURACIES ON THE SERVICE;
> 4. ANY INTERRUPTION OR CESSATION OF THE SERVICE;
> 7. THE REMOVAL OR UNAVAILABILITY OF ANY CONTENT.[158]

Since June 2022, 500 hours per minute were uploaded to YouTube.[159] Because the manual assessment of copyright infringement and exceptions or limitations of copyright is not scalable, platforms increasingly rely on automatic content recognition filters, such as YouTube's Content ID.

---

155 Danny Friedmann, *Sinking the Safe Harbour with the Legal Certainty of Strict Liability in Sight*, 9 J. INTELL. PROP. L. & PRAC. 148, 151 (2014); Danny Friedmann, *Oscillating from Safe Harbor to Liability, supra* note 69, at 291, 293.

156 DAVID J. HAND, DARK DATA: WHY WHAT YOU DON'T KNOW MATTERS 23 (2020) (distinguishing Dark Data Types 1 and 2, respectively: data we know are missing; and data we do not know are missing).

157 Terms of Service, YOUTUBE (effective Jan. 5, 2022) https://www.youtube.com/static?gl=HK&template=terms.

158 *Id.*

159 Lauri Ceci, *Hours of Video Uploaded to YouTube Every Minute as of February 2022*, STATISTA (Mar. 13, 2023), https://www.statista.com/statistics/259477/hours-of-video-uploaded-to-youtube-every-minute/.

According to YouTube's Copyright Transparency Report,[160] Content ID generated 757.9 million unique claims or copyright removal requests during the first halve of 2022. In 0.5 percent of the cases, the claim was disputed. This amounts to 3,690,786 disputed cases.[161] It is likely that other users did not know how to counterclaim, or had no time or were fearful of being sued.[162] Therefore, the number of false positives is probably higher, especially because automatic content recognition filters at the moment are not able to take exceptions and limitations and the three-step test in a meaningful way into account.[163]

The inner workings of machine learning, such as deep learning algorithms that are used for automatic content recognition filters, remain hidden. They are taking millions of data points as inputs and correlate specific data features to produce an output, which makes the creation of the results impossible to precisely explain and interpret. This black-box characteristic leads to a lack of algorithmic accountability.[164] This is amplified by what Cotter calls "black box gaslighting"—platforms that hide behind the inexplicability of the algorithm.[165] As mentioned above, this black-box characteristic is increased only in regard to de-ranking search results and recommendations and the decrease of traffic to the content. Moreover, there is no redress possible against these hidden measures.

### II. China's Regulation on Algorithmic Recommendations (RAR)

The Cyberspace Administration of China (CAC) promulgated the Provisions on the Management of Algorithmic Recommendations in Internet Information Services (Regulation on Algorithmic Recommendations (RAR)) on December 31, 2021, which came into effect on March 1, 2022.[166]

Article 1 RAR makes clear that the provisions are not only on the basis of the Cybersecurity Law, Data Security Law, Personal Information Protection Law, and Measures on the Administration of Internet Information Services, but also other relevant laws and administrative regulations. The Copyright Law of 2020,[167] Trademark Law of 2019,[168] and Anti-Unfair Competition Law[169] are clearly within the RAR's scope. Among other authorities, the State Administration for Market Regulation (SAMR) is also responsible for RAR,[170] just as it is for the China National Intellectual Property Administration (CNIPA).

[160] YouTube, Copyright Transparency Report H1 2022, https://storage.googleapis.com/transparencyreport/report-downloads/pdf-report-22_2022-1-1_2022-6-30_en_v1.pdf.

[161] *Id.* at 10.

[162] Timothy Geigner, *YouTube Copyright Transparency Report Shows The Absurd Volume Of Copyright Claims It Gets*, TechDirt (Dec. 8, 2021), https://www.techdirt.com/2021/12/08/youtube-copyright-transparency-report-shows-absurd-volume-copyright-claims-it-gets/.

[163] Friedmann, *supra* note 138, at 141.

[164] *See generally* Frank Pasquale, The Black Box Society: The Secret Algorithms That Control Money and Information (2015).

[165] Kelley Cotter, *"Shadowbanning Is Not a Thing":Black Box Gaslighting and the Power to Independently Know and Credibly Critique Algorithms,* 26 Info., Comm. & Soc'y 1226, 1227 (2023).

[166] RAR, *supra* note 122.

[167] Zhonghua Renmin Gongheguo Zhuzuoquan Fa (2020 Xiuzheng) (中华人民共和国著作权法(2020修正)) [Copyright Law of the People's Republic of China (2020 Amendment)] (promulgated by the Standing Comm. Nat'l People's Cong., Nov. 11, 2020, effective June 1, 2021), https://www.lawinfochina.com/display.aspx?lib=law&id=9280&CGid=.

[168] Zhonghua Renmin Gongheguo Shangbiao Fa (2019 Xiuzheng) (中华人民共和国商标法(2019修正) [Trademark Law of the People's Republic of China (2019 Amendment)] (promulgated by the Standing Comm. Nat'l People's Cong., Apr. 23, 2019, effective Nov. 1, 2019), https://www.lawinfochina.com/display.aspx?id=30316&lib=law&SearchKeyword=&SearchCKeyword=&EncodingName=gb2312.

[169] Zhonghua Renmin Gongheguo Fan Bu Zhengdang Jingzheng Fa (2019 Xiuzheng) (中华人民共和国反不正当竞争法(2019修正) [Anti-Unfair Competition Law of the People's Republic of China (2019 Amendment)] (promulgated by the Standing Comm. Nat'l People's Cong., Apr. 23, 2019, effective Apr. 23, 2019), https://lawinfochina.com/display.aspx?id=30315&lib=law.

[170] RAR, *supra* note 122, art. 3.

Article 1 RAR also clarifies the functions of the provisions, which include "protect[ing] the lawful rights and interests of citizens, legal persons, and other organizations, and promot[ing] the healthy and orderly development of internet information services." Article 6 RAR provides the same in negative form. Both provisions point to the need of recommendation algorithms to balance the rights and interests of users (citizens), platforms (legal persons), and copyright holders (legal persons and other organizations). Such a balance should help maintain a vibrant medium of expression while keeping intellectual property infringements to a minimum, and should also take copyright exceptions and limitations into account. Acting contrary to this would violate the freedom of expression as guaranteed under Article 35 Constitution of the PRC.[171]

The RAR prescribes transparent algorithms[172] and their optimization.[173] The problem is that automatic content recognition algorithms that are transparent are not able to take exceptions and limitations and the three-step test into account, and those that could, although that stage in the development of AI has not been reached yet or not yet implemented,are not transparent.

The transmission of unlawful information shall be immediately ceased once discovered, "and measures shall be employed to eliminate it or otherwise address it . . . ."[174] The latter leaves open the possibility of de-ranking suspected content. The need for the resources and the setting up of enforcement mechanisms of copyright infringements, but also taking limitations and three-step test into account, can be derived by implication from Article 6(2) RAR.

The algorithm should not push infringing content.[175] Even though Article 17 RAR provides users the possibility to opt out of a personalized recommendation algorithm, a general recommendation algorithm that prohibits or discourages infringing content prevails. One can argue that the platform should inform the user who searches for an alleged infringing work, that it has been removed and the reasons why.[176]

The platforms shall inform the users of which conduct is prohibited.[177] It would be logical that at least the uploader is informed that the traffic to some content is reduced and why, so that he or she can oppose this procedure, in accordance with RAR's requirement that algorithms need to have convenient and effective portals for complaints and appeals.[178]

The RAR is encouraging industrial organizations (platforms) to self-regulate.[179] But it seems that governmental supervision is necessary to enforce the RAR. Algorithms must be periodically checked,[180] if they are to be compatible with (or in conformity with) the RAR.

Algorithms need to set up convenient and effective portals for user appeals and public complaints or reports.[181]

### *III. The EU's Artificial Intelligence Act (AIA)*

On August 1, 2024, the EU's AIA came into effect.[182] It can be materially qualified as a product safety law addressed to those that use algorithms in the course of business, including platforms, to mitigate the risks for users and society at large, including copyright holders.

[171] XIANFA art. 35, § 1 (1982) (China).
[172] RAR, *supra* note 122, art. 4.
[173] *Id.* art. 4.
[174] *Id.* art. 9.
[175] *Id.* art. 10.
[176] *Id.* art. 17(3).
[177] *Id.* art. 16.
[178] *Id.* art. 22.
[179] *Id.* art. 5.
[180] *Id.* art. 8.
[181] *Id.* art. 22.
[182] AIA, *supra* note 133.

The proposed AIA uses a pyramid of risks, from unacceptable, to high, to low or minimal risk. Unacceptable risk, regulated by Title II, includes "AI systems used by public authorities, or on their behalf, for social scoring purposes."[183] In contrast to the prohibition under the AIA, in China, a connection between IP infringements and social score is allowed and used to deter dishonest behavior in regard to IP.[184]

High risk, regulated by Title III, is relevant to AI systems that create adverse impact on people's safety or their fundamental rights.[185]

One can argue that automatic content recognition filters should be qualified as high risk, because they may jeopardize fundamental rights such as the right of expression.[186] On the one hand, copyright in the EU has a catalogue of exceptions and limitations to copyright to promote this right of expression.[187] On the other hand, these exceptions and limitations are restrained to protect copyright holders.[188] Thus, automatic content recognition filters need to do a delicate balancing act.

So far, AI systems compare and identify uploaded works with works in a database that are allegedly eligible for copyright protection, and automatically either block those works or throttle the traffic to those works, without taking exceptions and limitations or fair use into account.[189] Despite significant numbers of false positives, automatic content recognition systems have not been qualified in the AIA as high risk. This contradicts with fundamental rights. Nevertheless, at the end of 2022, the Council of the EU called for promoting safe AI that respects fundamental rights,[190] without explicitly referring to the freedom of expression[191] or the protection of property or intellectual property.[192]

## G. Summary Shadow Banning

Just like generative search, shadow banning on social media platforms is opaque and reduces traffic to an affected website. Shadow banning involves algorithmic manipulation, such as de-ranking or reducing visibility of content, often to avoid legal risks or visibly and directly enforce copyright. These actions, though intended to protect rights holders, lack transparency and can chill free expression.

## H. Conclusion

The internet, once envisioned as a decentralized, democratized public sphere, is increasingly shaped by opaque practices like generative search and shadow banning, which undermine transparency, equity, and expression. Generative search diminishes website traffic by synthesizing content into direct answers, threatening the sustainability of independent content creators and small enterprises. Similarly, shadow banning on social media suppresses visibility and engagement, often through non-transparent algorithmic moderation aimed at avoiding legal risks or enforcing copyright. The public sphere, as conceptualized by Jürgen Habermas, runs the risk of being reduced to a simulation of dialogue structured by opaque commercial logic rather than democratic engagement.

[183]*Id.* art. 5.

[184]*The PRC Social Credit System and IP Protection*, SIPS (Apr. 4, 2019), https://sips.asia/knowledge/legislation-and-policy/the-prc-social-credit-system-and-ip-protection/.

[185]AIA, *supra* note 133, art. 6.

[186]Charter of Fundamental Rights of the European Union, art. 11, 2000 O.J. (C 364/1) [hereinafter CFREU] (entered into force Dec. 1, 2009); European Convention for the Protection of Human Rights and Fundamental Freedoms, art. 10, Nov. 4, 1950, 312 E.T.S. 5 [hereinafter ECHR].

[187]CFREU, *supra* note 186, art. 11.

[188]CFREU, *supra* note 186, art. 17.1; Protocol No. 1 to the European Convention on Human Rights art. 1, Mar. 20, 1952, E.T.S. 9 (establishing protection of property includes intellectual property). *See, e.g.*, Melnychuk v. Ukraine, App. No. 28743/03 (July 5, 2005), https://hudoc.echr.coe.int/eng?i=001-70089. *See generally* Guide on Article 1 of Protocol No. 1 to the European Convention on Human Rights: Protection of Property, EUR. CT. H.R. (Aug. 31, 2020), https://rm.coe.int/guide-art-1-protocol-1-eng/1680a20cdc [hereinafter Guide to Protocol No. 1].

[189]Friedmann, *supra* note 138, at 141.

[190]European Council Press Release 1008/22, Artificial Intelligence Act: Council Calls for Promoting Safe AI That Respects Fundamental Rights (Dec. 6, 2022), https://www.consilium.europa.eu/en/press/press-releases/2022/12/06/artificial-intelligence-act-council-calls-for-promoting-safe-ai-that-respects-fundamental-rights/.

[191]CFREU, *supra* note 186.

[192]CFREU, *supra* note 186, art. 17.1; Protocol No. 1, *supra* note 188, art. 1 (establishing protection of property includes intellectual property). *See, e.g.*, Melnychuk, App. No. 28743/03 at para. 3. *See generally* Guide to Protocol No. 1, *supra* note 188.

Website holders are often uncoordinated, and they have been too dependent on one search provider for their traffic referrals. Generative search, driven by technological progress for consumer convenience and reduced transaction costs, may evolve into a system where only large publishing houses benefit from licensing deals, further concentrating the news market. In this regard too, antitrust law has failed to fulfill its role. Government regulations should mandate that platforms provide a simple and accessible option for websites to opt out of having their information being used by generative search systems and provide strict conditions for generative search providers to avoid copyright infringement, plagiarism, and semi-plagiarism. Similarly, individual social media users are often connected solely through the network of a dominant platform. To address this systemic power imbalance, government regulation is needed to mandate transparency and the use of explainable algorithms (XAI).[193]

The EU and China both take a cautious path regarding algorithms in general and recommendation algorithms in particular, to safeguard their respective values. This is in stark contrast to the "[a]sk for forgiveness, not permission" approach, which might be conducive for technological innovation but might also lead to negative externalities. Paradoxically, this could mean that automatic content recognition filters that do take exceptions or limitations or fair use of copyright into account might be developed first in the U.S., if the pace of improvements of GPT-N and its ilk are any indication.

Legal and technological responses, such as China's RAR and the EU's AIA, attempt to address these challenges but remain insufficient in ensuring accountability, protecting fundamental rights, and fostering a fair ecosystem. Both practices reveal a troubling shift toward centralized control that prioritizes risk management and profit over open discourse, innovation, and the diversity of online expression. The stealthily applied algorithmic dissuasion via de-ranking and reduction in traffic of unauthorized but legal works can be qualified as extra-legal measures and is incompatible with the values of transparency and the fundamental rights of the users. In addition, one can argue that fundamental rights pre-empt licenses that are designed to immunize platforms against these measures.

To restore balance, platforms must adopt greater transparency and fairness in their algorithms, paired with robust redress mechanisms for affected users and creators. Licensing agreements, agentic AI, and clear regulatory frameworks can help mitigate harm, but only if designed to preserve the fundamental values of a democratic, decentralized internet. Without such measures, the modern "public square" risks becoming a tightly controlled and inaccessible space, further eroding its role as a platform for free expression and innovation.

**Acknowledgements.** The author declares none.

**Competing Interests.** The author declares none.

**Funding Statement.** No specific funding has been declared with respect to this Article.

[193] *GPT-4 is OpenAI's Most Advanced System, Producing Safer and More Useful Responses*, OpenAI, https://openai.com/product/gpt-4 (last visited Aug. 27, 2025).